\documentclass{iopart}
\usepackage{amssymb}
\usepackage{graphicx}

\newcommand{\dd}{\mathrm{d}}

\newcommand{\pd}[2]{\frac{\partial #1}{\partial #2}}

\newcommand{\mean}[1]{\langle #1 \rangle}
\newcommand{\Int}[1]{\int\dd #1\;}
\newcommand{\IInt}[3]{\int_{#2}^{#3}\dd #1\;}

\newcommand{\text}[1]{\mathrm{#1}}

\newcommand{\gam}{\gamma}

\newcommand{\kap}{\kappa}
\newcommand{\lam}{\lambda}

\newcommand{\sig}{\sigma}
\newcommand{\ra}{\rightarrow}

\newcommand{\kT}{k_\text{B}T}
\newcommand{\ts}{t_\text{s}}
\newcommand{\qs}{^\text{qs}}

\begin{document}

\title[Work distribution for the driven harmonic oscillator]{Work distribution
  for the driven harmonic oscillator with time-dependent strength: Exact
  solution and slow driving}
\author{Thomas Speck\footnote{Present address: Institut f\"ur
    Theoretische Physik II, Heinrich-Heine-Universit\"at, D-40225
    D\"usseldorf, Germany}}
\address{Department of Chemistry, University of California, and Chemical
  Sciences Division, Lawrence Berkeley National Laboratory, Berkeley,
  California 94720, USA}

\begin{abstract}
  We study the work distribution of a single particle moving in a harmonic
  oscillator with time-dependent strength. This simple system has a
  non-Gaussian work distribution with exponential tails. The time evolution of
  the corresponding moment generating function is given by two coupled
  ordinary differential equations that are solved numerically. Based on this
  result we study the behavior of the work distribution in the limit of slow
  but finite driving and show that it approaches a Gaussian distribution
  arbitrarily well.
\end{abstract}


\maketitle


\section{Introduction}

Driving a system away from thermal equilibrium requires work. Since thermal
fluctuations play a paramount role in small, mesoscopic systems it is only
consequent to also define a fluctuating work along single
trajectories~\cite{seki98,seif08}. While the celebrated
Jarzynksi~\cite{jarz97} and Crooks~\cite{croo99,croo00} non-equilibrium work
relations constrain the possible shapes, the actual distribution of the work
is still a non-universal function that depends on both the system dynamics and
the driving protocol. Analytical expressions for the work distribution in
isothermal processes are rather scarce except when the distribution is exactly
a Gaussian~\cite{spec05}. In particular, a single Brownian particle in a
moving harmonic potential has been studied extensively both
theoretically~\cite{mazo99,zon03,spec05,impa07b} and
experimentally~\cite{blic06,andr07} by trapping a colloidal particle with
optical tweezers.

The direct application of the Jarzynski relation to numerical or experimental
data in order to extract equilibrium free energy differences is marred by the
fact that it falls into the class of biased estimators. Rare, untypical
trajectories with low work values are exponentially weighted and a large
number of observations is required for estimates to converge. This number
typically grows exponentially with the mean dissipated work. Two principal
schemes have been discussed to address this problem: (i) reduction of the mean
dissipation by using optimal protocols~\cite{schm07} or escorted
simulations~\cite{vaik08}; (ii) unbiased estimators based on Bennett's
acceptance ratio method~\cite{coll05} or extended bridge
sampling~\cite{minh09}.

Of course, the easiest method to reduce the mean dissipated work might still
be to reduce the driving speed and keep the system close to equilibrium. In
this case a Gaussian has been predicted~\cite{hend01,spec04} as the limit
distribution for the work. However, as has been noted many times, a naive
application of a truncated Gaussian work distribution leads to wrong results
for the free energy difference (see, e.g., the discussion in
Ref.~\cite{croo07}). While the central limit theorem predicts that the center
of the distribution approaches a Gaussian this does not necessarily apply to
the extreme tails of the distribution, which are crucial for the correct
determination of free energy changes. For practical purposes it is, therefore,
important to understand the convergence of the work distribution towards its
limiting Gaussian distribution.

In this paper we study the probably simplest system that leads to a
non-Gaussian distribution of work: a single Brownian particle moving in a
one-dimensional, tightening harmonic potential. In addition to an exponential
tail for large work values the probability for negative work values is exactly
zero. Instead of determining the distribution of work directly we consider its
moment generating function and obtain a closed set of non-linear ordinary
differential equations. Using this exact result allows us to study the work
distribution as we reduce the driving speed.

\section{Work distribution and its generating function}

We consider a particle with position $x$ moving in a harmonic potential
\begin{equation}
  \label{eq:pot}
  U(x,t) = \frac{1}{2}k(t)x^2
\end{equation}
with time-dependent strength $k(t)$. Assuming overdamped dynamics the
evolution of the probability distribution $\rho_0(x,t)$ is well described by
the Fokker-Planck equation
\begin{equation}
  \label{eq:fp}
  \partial_t\rho_0 = L(t)\rho_0, \qquad
  L(t) \equiv \partial_x[k(t)x+\partial_x].
\end{equation}
Here and throughout the paper we measure energy in units of $\kT$ with
Boltzmann's constant $k_\text{B}$, length in units of $\sqrt{\kT/k_0}$, and
time in units of $\kT/(D_0k_0)$. The free diffusion coefficient of the
particle is $D_0$, and $k_0$ is the initial non-dimensionless value of the
trap strength (i.e., from now on $k(0)=1$). Given a driving protocol $k(t)$
from $t=0$ to $t=\ts$ the work is defined as the functional
\begin{equation}
  \label{eq:work}
  w[x(t)] = \IInt{t}{0}{\ts} \pd{U}{t}(x(t),t) 
  = \frac{1}{2}\IInt{t}{0}{\ts}\dot k(t)x^2(t)
\end{equation}
with switching time $\ts$. In particular, for $\dot k>0$ ($\dot k<0$) the work
is always non-negative (non-positive).

Since $x$ is the result of a stochastic process the work also will be a random
variable with distribution $p(w)$. For any driving protocol $k(t)$ the
Jarzynski relation
\begin{equation}
  \label{eq:jarz}
  \mean{e^{-w}} = \Int{w} e^{-w}p(w) = e^{-\Delta F}
\end{equation}
provides the link between non-equilibrium work and the equilibrium change of
free energy $\Delta F$, which for the system studied here becomes
\begin{equation}
  \label{eq:F}
  \Delta F \equiv F(\ts) - F(0) = \frac{1}{2}\ln k(\ts).
\end{equation}
For the analytical study of work distributions it has turned out to be
convenient to consider the joint probability $\rho(x,w,t)$ to find the system
in state $x$ at time $t$ and to have spent the accumulated work $w$ so
far. Its time evolution is governed by
\begin{equation}
  \label{eq:rho:fp}
  \pd{\rho}{t} = L\rho - \pd{U}{t}\pd{\rho}{w}
  = L\rho - \frac{1}{2}\dot kx^2\pd{\rho}{w}.
\end{equation}
In the following we work with the Laplace transform
\begin{equation}
  \label{eq:lap}
  \rho_\lam(x,t) \equiv \IInt{w}{0^-}{\infty} e^{-\lam w}\rho(x,w,t).
\end{equation}
For simplicity we focus on $\dot k>0$. After one integration by parts and
using $\rho(x,0^-,t)=0$ we obtain from (\ref{eq:rho:fp}) the evolution
equation
\begin{equation}
  \label{eq:sink}
  \partial_t \rho_\lam = L\rho_\lam - \frac{\lam\dot k}{2}x^2\rho_\lam.
\end{equation}
This is a linear reaction-diffusion type equation, called a 'sink' equation,
where diffusion is governed by the operator $L$ and 'probability' is created
or annihilated with a rate proportional to $\lam$, i.e., the function
$\rho_\lam$ is not normalized. Rather, its integral is the moment generating
function
\begin{equation}
  \label{eq:psi}
  \psi_\lam(t) \equiv \mean{e^{-\lam w}} = \Int{x} \rho_\lam(x,t).
\end{equation}
In particular, the mean and variance are obtained as
\begin{equation}
  \label{eq:meanvar}
  \mean{w} = -\mu_1, \qquad
  \sig_w^2 \equiv \mean{w^2}-\mean{w}^2 = \mu_2 - \mu_1^2,
\end{equation}
where $\mu_n\equiv\dd^n\psi_\lam/\dd\lam^n|_{\lam=0}$.

Already in one of the first experiments demonstrating a non-equilibrium work
relation~\cite{carb04} it has been noted that the distribution $p(w)$ for the
work is distinctly non-Gaussian. More recent computer simulations have found
exponential tails for large $w$~\cite{kwon11,nick11}. Indeed, considering the
extreme case of instantaneously switching from $k(0)=1$ to $k(\ts)=1+\Delta$
at some arbitrary time $0<t<\ts$ the work spent is $\Delta x^2(t)/2$. The work
distribution, therefore, becomes
\begin{equation}
  \label{eq:is}
  p^\infty(w) = \mean{\delta(w-\Delta x^2/2)} = (\pi w\Delta)^{-1/2} e^{-w/\Delta},
\end{equation}
where we have averaged over the equilibrium distribution
\begin{equation*}
  \rho_\text{eq}(x) = (2\pi)^{-1/2} e^{-\frac{1}{2}x^2}.
\end{equation*}
The generating function reads
\begin{equation}
  \label{eq:psi:is}
  \psi^\infty_\lam = \IInt{w}{0^-}{\infty} e^{-\lam w}p(w) 
  = (1+\Delta\lam)^{-1/2}.
\end{equation}
As a signature of the exponential tail $\psi^\infty_\lam$ diverges for
$\lam\ra\lam^\ast$ with $\lam^\ast=-\Delta^{-1}$.

The other limiting case is that of a quasi-static process with $\ts\ra\infty$
and work distribution $p\qs(w)=\delta(w-\Delta F)$, implying the generating
function
\begin{equation}
  \label{eq:psi:qs}
  \psi\qs_\lam(t) = e^{-\lam\Delta F} = e^{-\frac{\lam}{2}\ln k(t)} 
  = [k(t)]^{-\lam/2}.
\end{equation}
As a third exact result we note that for $\lam=1$
\begin{equation}
  \label{eq:hs}
  \rho_1(x,t) = (2\pi)^{-1/2} e^{-\frac{1}{2}k(t)x^2}, \qquad
  \psi_1(t) = e^{-\Delta F}
\end{equation}
solves the sink equation~(\ref{eq:sink}). The generating function $\psi_1$
obeys the Jarzynski relation, cf. (\ref{eq:psi}) with (\ref{eq:jarz}). This
particular solution to the sink equation~(\ref{eq:sink}) has been discussed
first by Hummer and Szabo~\cite{humm01}.


\section{Time evolution of the generating function}

From Eqs.~(\ref{eq:sink}) and~(\ref{eq:psi}) we obtain the equation of motion
\begin{equation*}
  \dot\psi_\lam = -\frac{\lam\dot k}{2}\phi_\lam
\end{equation*}
after integration over $x$, where
\begin{equation}
  \label{eq:phi}
  \phi_\lam(t) \equiv \Int{x} x^2\rho_\lam(x,t)
\end{equation}
is the generalized second moment. Multiplying (\ref{eq:sink}) by $x^2$ and
integrating again over $x$ results in
\begin{equation*}
  \dot\phi_\lam = -2k\phi_\lam + 2\psi_\lam
  - \frac{\lam\dot k}{2}\Int{x} x^4\rho_\lam.
\end{equation*}
By following this scheme we obtain a hierarchy of coupled ordinary
differential equations. Fortunately, for the harmonic oscillator a closure can
be found as follows. For $\lam=1$ we observe that the solution~(\ref{eq:hs})
is a, albeit not normalized, Gaussian. Since otherwise $\lam=1$ is not special
we conclude that $\rho_\lam$ is a Gaussian for all $\lam$. Then the closure
\begin{equation}
  \label{eq:closure}
  \Int{x} x^4\rho_\lam = \psi_\lam\Int{x} x^4\frac{\rho_\lam}{\psi_\lam}
  = \frac{3\phi_\lam^2}{\psi_\lam}
\end{equation}
follows from Gaussian statistics, expressing the generalized fourth moment in
terms of the functions $\psi_\lam$ and $\phi_\lam$. The resulting system of
non-linear, first-order ordinary differential equations
\begin{equation}
  \label{eq:ode}
  \dot\psi_\lam = -\frac{\lam\dot k}{2}\phi_\lam, \qquad
  \dot\phi_\lam = -2k\phi_\lam + 2\psi_\lam -  \frac{3\lam\dot k}{2}
  \frac{\phi_\lam^2}{\psi_\lam}
\end{equation}
is our first main result. Through inserting the Gaussian ansatz
\begin{equation*}
  \rho_\lam(x,t) = \sqrt{\frac{[\psi_\lam(t)]^3}{2\pi\phi_\lam(t)}}
  \exp\left\{ -\frac{x^2\psi_\lam(t)}{2\phi_\lam(t)} \right\}
\end{equation*}
into equation (\ref{eq:sink}) it is straightforward to check that
(\ref{eq:ode}) is indeed correct. Augmented by the initial conditions
$\psi_\lam(0)=\phi_\lam(0)=1$ these equations are readily solved numerically
by standard techniques.

\begin{figure}[t]
  \centering
  \includegraphics{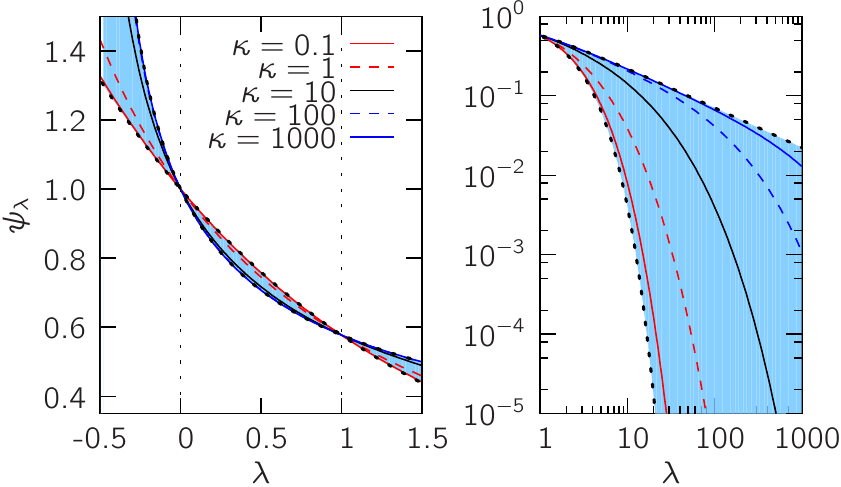}
  \caption{The generating function $\psi_\lam(t=\ts)$ \textit{vs}. $\lam$ for
    $\Delta=2$ and different driving speeds $\kap$. The shaded regions
    indicate the accessible range bounded by the
    quasi-static~(\ref{eq:psi:qs}) and instantaneous~(\ref{eq:psi:is})
    limiting cases (dashed lines). All lines cross at $\lam=0$ (normalization)
    and $\lam=1$ [Jarzynski relation~(\ref{eq:jarz})]. The right panel shows
    the decay of $\psi_\lam$ for large $\lam$.}
  \label{fig:psi}
\end{figure}

In figure~\ref{fig:psi} we plot the generating function $\psi_\lam(t=\ts)$ at
the final value of the control parameter for different driving speeds $\dot
k=\kap$ using the linear protocol
\begin{equation}
  \label{eq:lin}
  k(t) = 1+\kap t, \qquad \kap\ts = \Delta.
\end{equation}
The second relation fixes the switching time $\ts$. In the right panel of
figure~\ref{fig:psi} the generating function $\psi_\lam$ for large $\lam$ is
shown. The probability $p(0^+)$ to have spent no work is related to this
asymptotic behavior through the initial value theorem,
\begin{equation*}
  p(0^+) = \lim_{\lam\ra\infty}\lam\psi_\lam.
\end{equation*}
The numerical results suggest that even for large driving speeds $\psi_\lam$
decays faster than $\lam^{-1}$ with $p(0^+)=0$ and only for instantaneous
switching $p(0^+)\ra\infty$, see equation~(\ref{eq:is}). For large $w$ an
exponential tail is expected~\cite{nick11}, which corresponds to a divergence
of $\psi_\lam$ at $\lam^\ast<0$. Starting from $\psi_\lam(0)=1$ we expect a
singularity at time $t^\ast$ for all $\lam\leqslant\lam^\ast$, where
$\lam^\ast$ corresponds to $t^\ast=\ts$. Together with the results
Eqs.~(\ref{eq:psi:is}) and~(\ref{eq:psi:qs}) for the two limiting cases this
implies a value for $\lam^\ast$ that moves from $-\Delta^{-1}$ to $-\infty$
with decreasing driving speed $\kap$.


\section{Slow driving}

Of greater practical importance than quasi-static driving with $\kap\ra0$ are
processes with slow but finite driving speed $\kap$. In this case a Gaussian
distribution for the work has been predicted~\cite{spec04}. Of course, in the
present situation the work distribution can never be strictly a Gaussian since
$p(w\leqslant0)=0$ and because of the existence of an exponential tail for
large $w$.

\begin{figure}[t]
  \centering
  \includegraphics{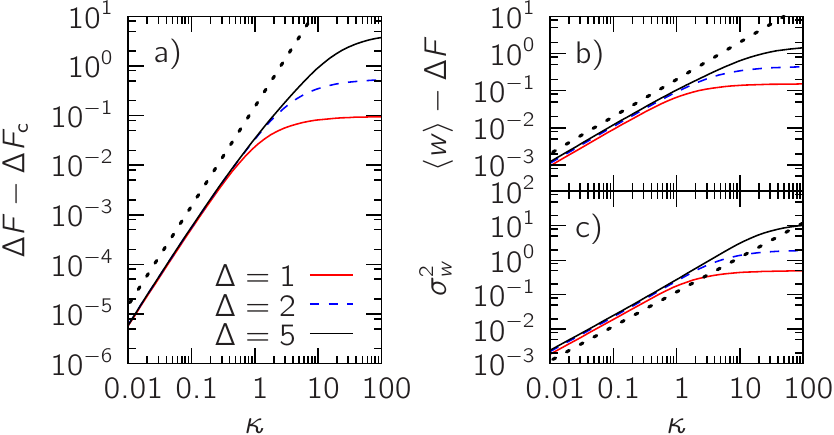}
  \caption{a)~Difference between true free energy $\Delta F$ and the truncated
    approximation (\ref{eq:Fc}) for the linear protocol (\ref{eq:lin}). The
    dashed line shows convergence with $\sim\kap^2$. Both mean work b) and
    variance c) grow linearly with driving speed (dashed lines $\sim\kap$).}
  \label{fig:slow}
\end{figure}

We calculate the true mean and variance [see (\ref{eq:meanvar})] by
integrating the four coupled equations
\begin{eqnarray*}
  \dot\mu_1 = -\frac{\kap}{2}\phi_0, \quad
  \dot\phi_0 = -2k\phi_0 + 2, \\
  \dot\mu_2 = -\kap\phi^{(1)}, \quad
  \dot\phi^{(1)} = -2k\phi^{(1)} + 2\mu_1 - \frac{3\kap}{2}\phi_0^2,
\end{eqnarray*}
which follow from (\ref{eq:ode}) after expanding
$\psi_\lam\approx1+\mu_1\lam+\mu_2\lam^2/2$ and
$\phi_\lam\approx\phi_0+\phi^{(1)}\lam$. Truncation of the cumulant expansion
for the free energy leads to the approximation
\begin{equation}
  \label{eq:Fc}
  \Delta F_\text{c} \equiv \mean{w} - \frac{1}{2}\sig_w^2.
\end{equation}
In figure~\ref{fig:slow}a) the difference $\Delta F-\Delta F_\text{c}$ is
plotted as a function of the driving speed $\kap$ for the linear protocol
(\ref{eq:lin}). It shows that the difference vanishes as $\kap^2$ and
that, therefore, the true free energy can be arbitrarily well approximated by
equation (\ref{eq:Fc}) through lowering the driving speed $\kap$. In
figure~\ref{fig:slow}b) and c) the mean dissipated work and variance are shown,
respectively.

\begin{figure}[t]
  \centering
  \includegraphics{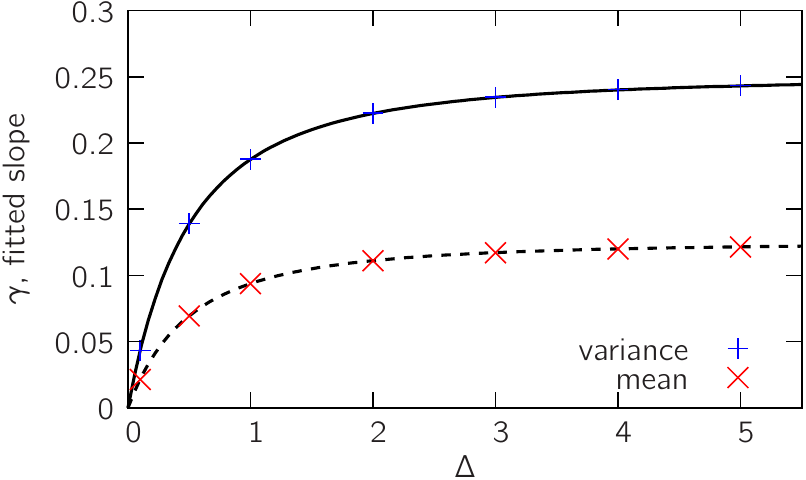}
  \caption{Comparison of the initial slope of mean [figure~\ref{fig:slow}b)]
    and variance [figure~\ref{fig:slow}c)] with
    equation~(\ref{eq:var:slow}). The solid line shows $\gam$ and the dashed
    line shows $\gam/2$.}
 \label{fig:slope}
\end{figure}

The strategy used in Ref.~\cite{spec04} to obtain the work distribution for
slow driving corresponds to expanding the generalized second moment
\begin{equation}
  \label{eq:phi:ex}
  \phi_\lam = \psi_\lam/k + \kap\phi_\lam^{(1)} + \cdots
\end{equation}
in powers of $\kap$. To first order we obtain from (\ref{eq:ode})
$\dot\psi_\lam=-\lam\kap \psi_\lam/(2k)$, which is solved by the quasi-static
solution $\psi\qs_\lam$ (\ref{eq:psi:qs}). For the next order we plug
the expansion (\ref{eq:phi:ex}) into (\ref{eq:ode}) and retain only
terms of order $\kap$ with $\dot\phi_\lam^{(1)}\sim\kap$. The result is
\begin{equation*}
  -(1+\lam/2)\frac{\psi_\lam}{k^2} = -2k\phi_\lam^{(1)} -
  \frac{3\lam}{2k^2}\psi_\lam
\end{equation*}
with
\begin{equation*}
  \phi_\lam^{(1)} = \frac{1-\lam}{2k^3}\psi_\lam.
\end{equation*}
The generating function now reads
\begin{equation}
  \label{eq:psi:slow}
  \psi_\lam(\ts) = \exp\left\{-\frac{\lam}{2}\ln
    k-\frac{\lam(1-\lam)}{2}\gam\kap \right\}.
\end{equation}
The corresponding distribution is of course a Gaussian with variance
$\sig_w^2=\gam\kap$ and mean $\mean{w}=\Delta F+\gam\kap/2$, where
\begin{equation}
  \label{eq:var:slow}
  \gam(\Delta) \equiv \IInt{t}{0}{\ts}\frac{\kap}{2(1+\kap t)^3} =
  \frac{\Delta(2+\Delta)}{4(1+\Delta)^2}.
\end{equation}
In contrast to the truncated Gaussian distribution obtained by using the true
mean and variance, the distribution following from (\ref{eq:psi:slow})
\emph{always} fulfills the Jarzynski relation~(\ref{eq:jarz}). In
figure~\ref{fig:slope} we compare the initial linear slope of mean and variance
for these two Gaussian distributions, where the slope of the numerical data
has been determined through a polynomial fit. The agreement between the two
Gaussian distributions is our second main result.

\section{Conclusions}

For a driven harmonic oscillator we have derived the equation of motion for
the moment generating function of the work distribution. Even though the exact
distribution is non-Gaussian it can be approximated by a Gaussian such that
the error vanishes with the square of the driving speed $\kap^2$. While such a
behavior is expected to hold in general~\cite{spec04} it remains to be
investigated to which extent the exponent 2 is system dependent. We have
focused on a linear protocol with $\dot k>0$ which corresponds to a stiffening
trap. The case $\dot k<0$ follows in analogy by defining the Laplace transform
in (\ref{eq:lap}) with respect to negative work values. However, it should be
kept in mind that \textit{slow} driving is defined with respect to the
time-scale separation between driving speed and the system's relaxation time
$1/k$, which for a widening trap increases strongly.

The extension to a time-dependent quadratic potential energy with many degrees
of freedom is straightforward through using the Wick theorem in
(\ref{eq:closure}). It will be worthwhile to study the method employed in this
paper for more complicated potentials $U(x)$. While in the present case the
Gaussian closure is exact, approximate closures for other potentials might
nevertheless lead to accurate results for the work probability and cumulants.


\ack

I thank Udo Seifert for many inspiring discussions and a critical reading of
the manuscript. Financial support by the Alexander-von-Humboldt foundation and
by the Director, Office of Science, Office of Basic Energy Sciences, Materials
Sciences and Engineering Division and Chemical Sciences, Geosciences, and
Biosciences Division of the U.S. Department of Energy under Contract
No.~DE-AC02-05CH11231 is gratefully acknowledged.


\end{document}